\documentclass[preprint,psfig]{emulateapj}

\usepackage{psfig}
\newcommand{\etal}{{et al.} }
\newcommand{\xmm}{{\it XMM-Newton} }
\newcommand{\xmmp}{{\it XMM-Newton}}
\newcommand{\chandra}{{\it Chandra} }

\newcommand{\hetg}{{\it HETGS} }
\newcommand{\hetgp}{{\it HETGS}}

\newcommand{\fekalfa}{{Fe~K$\alpha$} }

\newcommand{\fexxv}{Fe~{\sc xxv} }
\newcommand{\fexxvp}{Fe~{\sc xxv}}

\newcommand{\fexxvip}{Fe~{\sc xxvi}}

\newcommand{\feklyap}{{Fe~{\sc xxvi}~Ly$\alpha$}}

\newcommand{\loiv}{$L_{\rm [O IV]}$}

\newcommand{\hbeta}{${\rm H}_{\beta}$ }

\usepackage[dvips]{color}

\shorttitle{CHANDRA HEG OBSERVATIONS Fe K$\alpha$ LINES IN AGN}

\begin{document}

\title{ 
\chandra High Energy Grating Observations of the Fe K$\alpha$ Line Core in Type 2 Seyfert Galaxies: 
A Comparison with Type 1 Nuclei} 

\author{X. W. Shu\altaffilmark{1}, T. Yaqoob\altaffilmark{2}, J. X. Wang\altaffilmark{1}}

\altaffiltext{1}{
CAS Key Laboratory for Research in Galaxies and Cosmology, 
Department of Astronomy, University of Science and Technology of China, 
Hefei, Anhui 230026, P. R. China, xwshu@mail.ustc.edu.cn, jxw@ustc.edu.cn}
\altaffiltext{2}{Department of Physics and Astronomy,
Johns Hopkins University, Baltimore, MD 21218, yaqoob@pha.jhu.edu}

\begin{abstract}

We present a study of the core of the \fekalfa emission line at 
$\sim 6.4$~keV { in a sample of type II Seyfert galaxies 
observed by the \chandra High Energy Grating (HEG). 
The sample consists of 29 observations of 10 unique sources. 
We present measurements of the \fekalfa line parameters
with the highest spectral resolution currently
available. In particular, we derive the
most robust intrinsic line widths for some of the sources in the
sample to date.
We obtained a weighted mean FWHM of $2000 \pm 160 \ \rm km \ s^{-1}$ for 8
 out of 10 sources (the remaining sources had 
insufficient signal-to-noise). 
From a comparison with the optical emission-line widths { obtained from 
spectropolarimetric observations}, 
we found that the location of \fekalfa line-emitting material is a 
factor of $\sim0.7-11$ times the 
size of the optical BLR. 
Furthermore, compared to 13 type I AGNs for which the best \fekalfa line FWHM 
constraints were obtained, we found no difference in the FWHM distribution or the 
mean FWHM, and this conclusion is independent of the central black hole mass. 
This result suggests that
the bulk of the \fekalfa line emission may originate from a universal region 
at the same radius with respect to the gravitational radius, $\sim3\times10^4$ $r_g$ on average. 
By examining the correlation between the \fekalfa luminosity and the [O IV] line luminosity,
 { we found a marginal difference in the \fekalfa line flux between type I and type II AGNs, 
but the spread in the ratio of $L_{\rm Fe}$ to \loiv~ is about two orders of magnitude.  
Our results confirm the theoretical expectation that the
\fekalfa emission-line
luminosity cannot trivially be used as a proxy of the intrinsic
AGN luminosity, unless a detailed comparison of the data with proper models is applied.}}
\end{abstract}


\section{INTRODUCTION}
\label{intro}
Both type I and type II active galactic nuclei (AGN) are known to exhibit a 
narrow (FWHM~$<10,000 \ \rm km \ s^{-1}$) 
 \fekalfa fluorescent emission line at $\sim 6.4$~keV in their X-ray spectrum  
(e.g. Sulentic \etal 1998; Lubi\'{n}ski \& Zdziarski 2001;
Weaver, Gelbord, \& Yaqoob 2001; 
Perola \etal 2002; Yaqoob \& Padmanabhan 2004 (hereafter YP04); 
Levenson \etal 2002, 2006; { Jiang et al. 2006;}
 Winter \etal 2009; Shu, Yaqoob \& Wang 2010, hereafter paper I). 
The line profile of the \fekalfa core
is important for probing its origin and it can provide unique information on the
dynamics and physical state of the line-emitting region
(Yaqoob et al. 2001, 2003, 2007).
While it is widely accepted that the narrow \fekalfa line cores are produced in cold, neutral 
matter far from the nucleus, the exact location and distribution of the line-emitting gas 
still remain uncertain (see Paper I, and references therein). 
Nandra (2006) first examined the relation between the \fekalfa and optical H$\beta$ line widths, 
which can potentially give a direct indication of the location of
 the \fekalfa line-emitting region relative to the optical 
broad-line region (BLR). However, 
the results were ambiguous, and the data allow for an origin of the \fekalfa line anywhere from the outer 
regions of an accretion disk, the BLR, and a parsec-scale torus.
Meanwhile, Bianchi et al. (2008) reported a meaningful comparison between the \fekalfa and 
H$\beta$ line widths in NGC 7213, and found the FWHM of both lines 
are consistent with each other ($\sim$2500 km s$^{-1}$), implying a BLR origin of the \fekalfa
emission line.
Using the high-energy grating (HEG)
on the \chandra \hetg ({\it High Energy Transmission Grating Spectrometer},
see Markert et al. 1995), which affords the
best spectral resolution currently available in the Fe K band
(at 6.4 keV is $\sim$39 eV,
or $\sim$1860 km s$^{-1}$ FWHM), in Paper I we  
presented a more thorough and comprehensive study of the \fekalfa line core emission in a large sample of 
type I Seyfert galaxies (see also YP04). 
We measured the intrinsic width of the narrow \fekalfa line core and obtained a weighted mean of FWHM = 2060 $\pm$ 
230 km s$^{-1}$.
A comparison with the optical emission-line widths suggested that there may not 
be a universal location of the \fekalfa line-emitting region relative to the BLR, consistent with the results of Nandra (2006).  Our results in fact showed that the location of the \fekalfa line emitter relative the BLR 
appears to be different from source to source.

However, one must be cautious in explaining the origin of the narrow \fekalfa core, 
especially for measurements made with instruments that have lower spectral
resolution than the \chandra \hetg (e.g., see Liu et al. 2010 for NGC 5548).
The uncertainty in the intrinsic line width measurements is usually large, and part of the narrow component 
may have a contribution from an underlying broad line in some (if not all) unobscured AGNs 
(e.g., Lee et al. 2002; YP04; Nandra 2006).
 However, in cases when the narrow \fekalfa core is unresolved even with the 
\chandra HEG, such contamination is not an issue, and from 
the upper limits on the line widths we can place strong constrains on the origin of the 
intrinsically narrow \fekalfa line core (see Paper I). 
In the paradigm of the unification model (Antonucci 1993), the narrow \fekalfa line 
emission in type II AGNs is expected to be produced in the same material as in type I nuclei. 
Therefore, one of the things that we would like to know is whether there is any systematic difference in 
the origin of the \fekalfa line in type I and type II AGNs. 
In addition, 
for a given geometry and model,
one would expect a particular relation between the equivalent width (EW) of 
the \fekalfa line, $N_H$, and the orientation of line-emitting structure
(e.g., Murphy \& Yaqoob 2009; Ikeda, Teramshima, \& Awaki 2009;
Yaqoob et al. 2010; Brightman \& Nandra 2011). Thus a comparison of \fekalfa 
emission-line luminosities and EWs in
type I AGNs versus type II AGNs can potentially offer basic tests of some
of the predictions of AGN unification scenarios.

In this paper, we present a study of the narrow \fekalfa emission line in 
a sample of type II AGNs observed with the Chandra \hetgp.
The observations and spectral fitting are described in Section 2. In 
Section 3, we present the results for the properties of the core of 
the narrow \fekalfa emission line and their implications. 
In Section 4, we summarize our results and findings.
Throughout this paper, we adopt a cosmology of $\Omega_M=0.3, \Omega_{\lambda}=0.7$, 
and H$_0=70$ km s$^{-1}$ Mpc$^{-1}$.     
 
\section{OBSERVATIONS AND SPECTRAL FITTING}
\label{data}
{  
Our study is based on {\it Chandra} {\it HETGS} observations of 
AGNs, which were in the {\it Chandra} public archive as of 2010 Aug 1. 
In Paper I, we presented a study of the properties of 
the \fekalfa line emission in 
 a sample of predominantly type I AGNs. In this paper, 
we expand the sample of AGN in Paper I to include type II sources 
by simple modifications
of the selection criteria. One is to lift any constraints on the
column density (i.e., there now no selection on column density), and the
other is to lower the acceptance 
threshold for the total number of counts in the HEG
spectra. The sample in Paper I was composed of the least
complex spectra due to a maximum column density criterion,
and a threshold of a total number of HEG counts of 1,500.
The reader is referred to Paper I for the details of the remaining
selection criteria, which were not changed.  
Lowering the signal-to-noise
threshold was necessary because the larger equivalent widths
of the Fe~K$\alpha$ line for heavily absorbed sources result
in useful constraints for sources that are weaker than those
with smaller equivalent widths. 
With the two modified
selection criteria, we end up with a sample that now includes
both type I and type II AGN, as well as intermediate types.
The present paper reports on the results of an analysis of 
29 observations of the 10
sources in the new, larger sample that were not included in Paper I.
Not surprisingly, the sources in the present paper are all classified
as type II in NED (including NGC 1275, also known as 3C~84 and
Perseus A, which lies at the center of a rich cluster of galaxies,
and has a peculiar optical emission-line spectrum with two
distinct narrow emission-line systems, e.g., Conselice \etal 2001).

We note that the relationship
between the optical classification of an AGN as type I to type II 
and the X-ray absorption properties is not always clear cut,
so the sample in Paper I included some AGN that were not
type I. In particular, Paper I included NGC~5506, MCG~$-$5-23-16,
and NGC~2110, which are type 1.9 AGN. 
When comparing the results for the type II AGN in the present paper,
we simply refer to the intermediate Seyfert types in Paper I as
type I for convenience (this is more of a reflection of the relative
simplicity of the X-ray spectrum).
We note that since the publication of Paper I, 15 new observations
of two of the AGN in Paper I (NGC~4051 and Akn~564) have been made public.
The new observations have been analyzed and this information has been
incoorporated into the comparisons between the results in Paper I and the
present paper.
 
}
The \chandra \hetg consists of two grating assemblies,
a High-Energy Grating (HEG) and a Medium-Energy Grating (MEG),
and it is the HEG that achieves the highest spectral resolution.
The MEG has only half of the spectral resolution
of the HEG and less effective area in the Fe~K band, so our study will
focus on the HEG data. 
Out of the 10 unique sources, 8 were observed more than once: namely NGC 4945, 
the Circinus galaxy, NGC 6240, NGC 1068, Centaurus A, NGC 4388, NGC 4258 and NGC 1275.   
The remaining two sources that were observed only once are NGC~4057 and Mrk~3.
In order to obtain the highest precision measurements of the 
\fekalfa emission-line parameters, in particular for the width 
of the line, we concentrate on the analysis of 
the co-added spectra for the sources with multiple observations. 
{ We checked that for sources observed more than once, the count rates variations are 
very small (less than 10\%). Therefore, in this study we report only the results of 
time-averaged spectroscopy.}
The mean HEG counts rates ranged from $0.006 \pm 0.001$ ct/s for the weakest source (NGC 4945) to 
$0.792 \pm 0.004$ ct/s for the brightest source (Centaurus A). The exposure time ranged from 
$\sim 10$~ks 
to $\sim 170$~ks per observation, and the largest net exposure time from summed data from
observations of the same source was $\sim 440$~ks (NGC~1068). 
{ Details of the 29 observations used in this paper are given in Table 1.}

The \chandra data for the sample 
were reduced and HEG spectra were made 
as described in Yaqoob \etal (2003) and YP04.
We used only the first orders of the grating data (combining
the positive and negative arms). 
Further details of all of the observations can be found
in the \chandra data archive
at {\tt http://cda.harvard.edu/chaser/}. Higher-level products,
including lightcurves and spectra for each observation
can be found in the databases {\it HotGAS} 
({\tt http://hotgas.pha.jhu.edu}), and {\it TGCat}
({\tt http://tgcat.mit.edu/}).
Background was not subtracted since it is negligible over the
energy range of interest (e.g. see Yaqoob \etal 2003).
Note that 
the systematic uncertainty in the HEG
wavelength scale 
is $\sim 433 \ \rm km \ s^{-1}$ ($\sim 11$~eV) at 6.4 keV
\footnote{http://space.mit.edu/CXC/calib/hetgcal.html}. 

The spectra were analyzed
using the spectral-fitting package XSPEC (Arnaud 1996).
Since we are interested in utilizing the highest possible
spectral resolution available, we used spectra binned
at $0.0025\AA$, and this amply oversamples the HEG resolution
($0.012\AA$ FWHM).
The $C$-statistic was used for minimization.
All model parameters will be
referred to the source frame.
Our method is simply to fit a simple absorbed continuum plus Gaussian
emission-line model over the 3--10~keV band for each spectrum, 
to avoid the influence of the complex absorption and many soft X-ray 
emission lines below 3 keV (e.g., Levenson et al. 2006), 
while providing enough high-energy coverage to fit the continuum. 
Galactic absorption was not included for any of the sources because
such small column densities have little effect above 3~keV.
For NGC 4258 and NGC 1275, the \fekalfa line was 
only weakly detected: $C$ decreased by 3.4 and 2.0, respectively, 
when a narrow, unresolved emission line at 6.4 keV was added to the
continuum. These figures correspond
to detections at less than 95\% confidence, and less than 90\% confidence
for NGC 4258 and NGC 1275 respectively (for the addition of one free parameter).
For these two sources we were therefore
only able to obtain constraints on the line flux 
(and not on the line centroid energy and line width), so the 
Gaussian model had one free parameter. 
Note that a significant \fekalfa emission line has been detected in NGC 4258 by 
$Suzaku$ and \xmm (Yamada et al. 2009; Reynolds et al. 2009), with EW=$45\pm17$ eV 
and EW=$53\pm19$ eV, respectively, which is consistent with the upper limit of 32 eV 
(at 90\% confidence) obtained by the \chandra HEG.

Thus, except for the two cases mentioned above, for the remaining spectra the Gaussian 
model component 
had three free parameters and there were a total of six free parameters in the
model, namely the power-law continuum slope ($\Gamma$), the overall normalization
of the power-law continuum, the column
density, $N_{H}$, the centroid energy of the Gaussian
emission-line component, $E_{0}$, its flux, $I_{\rm Fe~K}$, and
its width, $\sigma_{\rm Fe~K}$.
The approach of using an oversimplified
continuum model is necessitated by the limited bandpass
of the HEG data ($\sim 3-10$~keV) but since we are
interested in the narrow core of the \fekalfa emission
line, at the spectral resolution of the HEG, this is
not restrictive. 
As we will show in Section 3.1, in cases when the \fekalfa line 
was significantly detected, its centroid energy as well as its intrinsic width 
can be well-measured by the \chandra HEG.  
 We note that even for cases where we can obtain a reliable measure of the \fekalfa line FWHM, 
 the true line width may be less than the FWHM deduced from our 
simplistic model-fitting because there may be blending from an unresolved 
Compton-shoulder component { (e.g., Matt 2002; Yaqoob \& Murphy 2011).} 
However, this artificial broadening is not an issue for the 
line parameters of the high signal-to-nosie ratio HEG observations reported here. 
For example, although Circinus has one of the most robust detections of a 
Compton-shoulder component { (e.g., Bianchi et al. 2002), 
it has the smallest FWHM measure among the HEG sample.}
Obviously, use of such an empirical model
means that we should not assign a physical meaning to
$\Gamma$ and $N_{H}$.
We emphasize that our approach in the present paper is to perform
a very simple {\it empirical} analysis in order to obtain robust
measurements of the basic narrow \fekalfa line core
parameters that are not dependent on details of how the continuum is modeled. 
A realistic physical model of the continuum and the analysis of the 
soft X-ray spectral components is
 beyond the scope of this work and will be presented elsewhere.

\section{RESULTS AND DISCUSSIONS}
\subsection{Spectroscopy in the Fe K Band}
\label{hegspec}
Figure 1 shows the HEG spectrum for each source in the Fe K region, 
corrected for instrumental efficiency
and cosmological redshift. The spectra are binned at 0.01 \AA , similar to the HEG 
FWHM
spectral resolution of 0.012 \AA. The statistical 
uncertainties correspond to 68\% confidence Poisson errors, which we calculated using 
equations (7) and (14) in Geherls (1986) that approximate the upper and lower errors
 respectively.  
For sources which were observed more than once, the time-averaged spectra are shown. 
It can be seen that in some spectra,
emission lines from different ionized species of Fe are evident, 
in addition to the
narrow \fekalfa line at $\sim$6.4 keV, but the latter is always the strongest line.
Although in the present paper we are concerned only with the \fekalfa line core centered at $\sim$
6.4 keV, overlaid on 
the spectra in Fig. 1 are vertical dashed lines marking the 
expected positions of the \fexxv He-like 
triplet lines (the two intercombination lines are shown separately), \fexxvip~Ly$\alpha$, Fe I K$\beta$, 
and the neutral Fe K-shell threshold absorption-edge energy. The values adopted for these energies 
were from NIST2 (He-like triplet); Pike et al. 1996 (\fexxvip~ Ly$\alpha$); Palmeri et al. 2003 
(Fe I K$\beta$ ), and Verner et al. 1996 (Fe K edge). 
 We emphasize that the emission from higher ionization states of Fe has little effect on the 
measurements of the intrinsic width of the \fekalfa line.  
For example, for NGC 1068 the spectral plot in the Fe K band shows a broad emission feature 
on the blue side of the \fekalfa line peak,
which is probably due to multiple species of ionic Fe. 
{The occurrence of highly ionized Fe emission lines in AGN X-ray spectra 
has been noted since the $ASCA$ observations, and they are extensively studied recently with 
\xmm and $Suzaku$ (e.g., Bianchi et al. 2005; Fukazawa et al. 2011). }
However, we can confirm from the 99\% confidence contour of \fekalfa line intensity 
versus FWHM (Figure 2), that the \fekalfa line 
width is still well-constrained (less than $\sim$3500 km s$^{-1}$).
This conclusion is also
consistent with the physical width of the narrow core 
of the \fekalfa line in the spectral plot that is apparent simply from visual inspection.

The best-fitting \fekalfa emission-line parameters for each spectrum 
are shown in Table 2. 
We do not give the best-fitting values of $\Gamma$ or $N_{H}$
in Table 2 because the values derived using the simplistic
continuum model are not physically meaningful but are
simply parameterizations.
We obtained a weighted mean line-center energy of 6.397$\pm$0.001 keV for the 24 observations 
of 8 sources. 
At 99\% confidence (for three free parameters), none of the AGNs has a line centroid energy 
greater than $\sim$6.41 keV, 
indicating that the line-emitting matter is cold and 
essentially neutral, consistent with results presented in Paper I for a sample of type I AGNs. 
Here, and hereafter, for the calculation of the weighted mean of
any quantity with asymmetric errors, we
simply assumed symmetric errors, using the largest 68\% confidence
error from spectral fitting.

\subsection{The Intrinsic Width of the \fekalfa Line Emission}

From our spectral fits (Table 2) and the joint confidence contours of \fekalfa line intensity 
versus FWHM (Figure 2 and 3), we found that at 99\% confidence, 
the \chandra HEG resolves the narrow component of the \fekalfa emission in 8 out of 10 sources. 
The weighted mean FWHM of the \fekalfa line cores is $2000 \pm 160 \ \rm 
km \ s^{-1}$.  
{ 
The fact that this mean is approximately equal to the
HEG FWHM spectral resolution is not indicative of a calibration
bias. This is supported by the fact that the FWHM of the Fe~K$\alpha$ line
in Circinus is {\it less} than the HEG FWHM at a confidence level of
greater than 99\%. In addition, the spectral resolution for gratings
does not degrade with time because it is determined principally by
spatial dispersion, and the resolution is well-established from
bright Galactic sources with narrow lines 
(http://space.mit.edu/CXC/calib/hetgcal.html).
Note that the weighted mean value is consistent with the straight mean of
the FWHM, of $2510 \pm 160 \ \rm km \ s^{-1}$.
}
The measurements of the intrinsic width of the line can potentially be used to constrain 
the location of the medium responsible for the core of the \fekalfa emission line 
(e.g., Yaqoob et al. 2001). 
Comparing the \fekalfa line
FWHM with that of the optical broad emission lines (e.g., H$\alpha$ and/or \hbeta lines) can 
give a direct indication of the location of the \fekalfa
line-emitting region relative to the optical broad-line region (Nandra 2006, 
Bianchi et al. 2008, Paper I). 
To place limits on the location of the \fekalfa emitter relative to the BLR, 
we compiled from the literature 
the width of optical broad emission line from spectro-polarimetric observations for 
five sources in our sample, and the values of the H$\alpha$ or \hbeta FWHM 
are listed in Table 2.  
{
The near-infrared broad lines have not been taken into account in the
present paper, as they are possibly produced in a region that differs
from the BLR in density, and/or in the amount of extinction
(e.g., Ramos Almeida \etal 2008; Landt \etal 2008). However, in
cases where there is no spectropolarimetric data at all, the FWHM
of near-infrared lines might still yield some useful information}\footnote{ We searched in the literature and found that only one source, NGC 6240, 
has NIR broad-line Br$\alpha$ measurement reported, with FWHM=1800$\pm$200 km/s (Cai et al. 2010). 
This value is consistent with the width of the Fe K line at 99\% confidence.}.
 
Figure 3 shows the 99\% confidence contours of
the \fekalfa line intensity versus the {\it ratio}
of the \fekalfa FWHM to optical line FWHM. We found that at
the two-parameter 99\% confidence level, the FWHM ratio lies in the range of $\sim0.3-1.2$.
 At 99\% confidence, the \fekalfa line widths in 3 out of 5 sources are less than that 
of optical lines, suggesting that the observed \fekalfa emission is not likely produced  
in a BLR where the optical lines are produced. The remaining two sources ({ NGC 4388 and NGC 1068}) 
could have consistent 
width for both lines. 
This would be expected in at least one source, NGC 4388, which is known 
to have significant column density variations (Elvis et al. 2004), if its BLR 
acts as absorber and emitter seen in the X-ray band.    
In fact, there is now significant evidence that the BLR may act as an X-ray absorber, 
like in the cases of NGC 1365 and NGC 7582 (e.g., Risaliti et al. 2005; Bianchi et al. 2009). 
{ However, if the BLR is obscured by a fully covering column density greater than $\sim 5 \times
10^{23} \ \rm cm^{-2}$ ({ as in the cases of NGC 4507, Mrk 3 and the Circinus} ) any Fe~K$\alpha$ line component from the BLR will be greatly diminished,
or undetectable. If the BLR is obscured by such a large column density, the only way it would
be possible to observe an Fe~K$\alpha$ line component from the BLR is if the BLR is
not fully covered.}
{ 
The fact that the \fekalfa line width of Compton-thick source NGC 1068 is consistent with that 
of the polarized \hbeta line (within 99\% errors) could be simply attributed to the large 
error in the \fekalfa line width. Alternatively, this could indicate that the 
heavy obscuration which blocks the BLR has comparable size with that of the BLR. 
}


Since Keplerian velocities are inversely proportional to the square root of the
orbit size, { the observed distribution of the ratio of the \fekalfa to optical line FWHM} 
implies that the \fekalfa line-emitting region size could be   
a factor $\sim 0.7-11$ times larger than the optical line-emitting region. 
We obtained a weighted mean of 0.57$\pm$0.05 for the ratios of the 
\fekalfa line FWHM to the optical line FWHM, 
corresponding to the \fekalfa line-emitting region being, on average, $\sim$3 times the 
size of BLR. However, 
for AGNs in general, there may be no universal location of the \fekalfa line-emitting region within 
 a factor of $\sim$1--10 times size of BLR. 
However, as we will show later, the \fekalfa line emitter may be associated with a universal location in 
terms of
gravitational radius ($r_g$, where $r_g=GM_{\rm BH}/c^2$).  
If the \fekalfa line emission has a significant contribution from the putative obscuring torus 
that is required by AGN unification models, 
our results show that the size-scale of the torus may be smaller
than traditionally thought. Note that 
Gaskell, Goosmann, \& Klimek (2008) argue that there is considerable 
observational evidence that the BLR itself has a toroidal structure, 
and that there may be no distinct boundary between the
BLR and the classical parsec-scale torus. 

One of the interesting things that we can investigate
is whether the material responsible for producing the 
narrow \fekalfa 
line in type II AGNs systematically differs from that in type I objects. 
Of the AGNs in the sample reported in Paper I, 
we identified { thirteen} type I AGNs 
(including 3 moderately obscured sources with 
weak broad Balmer lines, formally classified as intermediate type AGNs) that provided the very best 
statistical constraints on the \fekalfa line FWHM
(as indicated by the two-parameter confidence contours of line flux versus 
FWHM; see Figure 6 in Paper I). 
The weighted mean FWHM of the \fekalfa core { is 2170$\pm$220 km $^{-1}$}, in 
good agreement with the measurement for type II AGNs in the
present sample. 
{ However, we found a slightly larger FWHM of the \fekalfa line (3620$\pm$220 km $^{-1}$) for 
type I AGNs if the straight mean is used, but as we will show below, it
may be due to a bias towards the measurements with lower quality data.}
We show in Figure 4 the distribution of \fekalfa line FWHM for the current sample 
(solid line), compared with the { thirteen} type I AGNs in Paper I (dashed line). 
It can be seen that both histograms are not Gaussian and are not strongly peaked. 
In addition, we found that the measurement of the FWHM in one AGN (namely MCG -6-30-15) deviated significantly 
from the distribution of the rest of the sources.   
From our empirical analysis, we obtained FWHM=11880$^{+4650}_{-4030}$ km s$^{-1}$ (see Paper I) 
for this object, and the corresponding 
90\% confidence lower limit (for three free parameters) is $\sim$6000 km s$^{-1}$. Note that 
 MCG -6-30-15 has the strongest and broadest \fekalfa line yet 
observed in an AGN. The larger FWHM obtained from our empirical analysis could be attributed to the \fekalfa core from 
underlying disk-line component, and/or a complex continuum (e.g., Miller et al. 2008).  
Note that Young et al. (2005) analyzed the \chandra HEG spectrum by using more complex models 
including both disk-line and narrow \fekalfa core emission, and obtained an FWHM$<$ 4700 km s$^{-1}$ for the 
narrow component, which is consistent with those from the other sources in our sample. 
However, as the narrow \fekalfa line EW is relatively low ($\sim$60 eV, see Paper I), we cannot tell 
whether the line profile is affected by the complex continuum, or 
whether it is intrinsically broader than the other AGNs
. Future X-ray missions, such as 
{\it Astro--H}, which has much higher spectral resolution, will help to measure the true profile of the 
narrow \fekalfa emission line in this object.  
{ Even considering }the one AGN mentioned above, 
there is not a significant difference in the distribution of FWHM for the two subsamples. 
A Kolmogorov-Smirnov (K-S) test shows that the probability 
that the type I ({ including MCG -6-30-15}) and type II AGNs are drawn from the same parent population is 0.83.
Therefore, it appears that there is no difference in the origin of the 
\fekalfa line in type I and type II AGNs, which is consistent with the predictions of 
the unification model. 

Figure 5 shows the relationship between the \fekalfa line FWHM and the black hole mass
(see Table~2 caption for references). 
Open and solid circles denote type I and type II AGNs, respectively. 
We see that all FWHM values, within the statistical errors,
 are consistent with a constant, independent of the mass of black hole. 
Assuming that the line originates in material that is in a virialized orbit around the black hole, we can 
estimate the distance, $r$, of the line-emitting material to the black hole using the relation $GM_{\rm BH}=
r\langle v^2\rangle$. Assuming 
that the velocity dispersion is related to FWHM velocity as $\langle v^2\rangle$=$3\over4$$v^2_{\rm FWHM}$ 
(Netzer et al. 1990), we can obtain 
$r=(4c^{2})/(3v^{2}_{\rm FWHM})r_g$, where $r_g=GM/c^2$. 
Using the weighted mean of FWHM $\sim$2000 km s$^{-1}$, we find that
the $r\sim3\times10^4r_g$, larger than the typical 
size of the BLR (e.g., Peterson et al. 2004).  
Therefore, our results seem to support the bulk of the \fekalfa line production arising from a region 
that is appears to be located at
a universal distance with respect to the gravitational radius, which is controlled by central black hole mass.
Note that Nandra (2006) also examined the relation between the FWHM of \fekalfa line and the black hole mass, 
but their results were ambiguous.  
The reason why they did not find the result that we found is possibly due to the
fact that some of the sources in their sample had only poor
quality and/or problematic \fekalfa line-width measurements. Including such sources
could have obscured the underlying result that $r/r_{g}$ may be the key
factor that determines the location of the line-emitting region.  
However, it is important to note that even with superb \chandra HEG spectral resolution, there may be blending from 
a Compton-shoulder component and/or 
multiple low ionization states of Fe (e.g., Yaqoob \& Murphy 2011), 
so that the true line width may be less than the FWHM deduced from 
our simplistic model-fitting. 

\subsection{What can we learn from the \fekalfa Line Emission? } 
From a theoretical point of view, the flux or luminosity 
 of the \fekalfa line ($L_{\rm Fe}$) is 
nontrivial to calculate, as it depends on a number of factors, including geometry, orientation, 
covering factor, element abundance, and column density of the line-emitting material. 
Using Monte Carlo simulations of a toroidal X-ray reprocessor model, Murphy \& Yaqoob (2009, 
see also Yaqoob et al. 2010) showed that geometrical and inclination-angle effects become 
important for $N_H\gtrsim10^{23}$ cm$^{-2}$ for the observed EW and line flux. 
For a
given covering factor and set of element abundances, a toroidal structure with
a column density of  greater than 
$\sim 5 \times 10^{24}$ cm$^{-2}$, observed at an edge-on inclination could produce 
an \fekalfa emission line flux that is an order of magnitude or more weaker than 
a face-on orientation (Yaqoob et al. 2010). 
In particular, the \fekalfa line luminosity is not simply a linear function of $N_H$, and it 
has a maximum value for a column density in the range $\sim3-8\times10^{23}$ cm$^{-2}$, 
depending on the inclination angle.
 Therefore, the \fekalfa line luminosity cannot be trivially used to measure the intrinsic AGN luminosity 
($L_{\rm AGN}$). The relation between $L_{\rm Fe}$ and $L_{\rm AGN}$ is more complicated 
than we would expect from optically-thin matter, 
and it is strongly model dependent,
and in particular it depends on covering factor, inclination angle and
column density. 
We emphasize that the column density for producing the 
\fekalfa line does not refer to the line-of-sight value, 
but rather to a value that corresponds to
the angle-averaged flux over all incident X-ray continuum radiation 
(see Murphy \& Yaqoob 2009).

Recently, Liu \& Wang (2010) confirmed that $L_{\rm Fe}$ and $L_{\rm AGN}$ are not simply related. 
They compared the narrow \fekalfa line emission between type I and type II AGNs, using 
data obtained with \xmmp, and found that statistically,  { the \fekalfa line luminosities in 
Compton-thin and Compton thick type II AGNs are about 2.7 and 5.6 times lower than that 
in type I sources, respectively. 
 They therefore proposed that different
 correction factors should be applied if one uses the \fekalfa
 line emission to estimate the AGN's intrinsic luminosity.}
{ To examine the relation between $L_{\rm Fe}$ and $L_{\rm AGN}$ for
 our \chandra grating sample, 
we plot in Figure 6 (upper panel) 
the \fekalfa line luminosity ($L_{\rm Fe}$), versus 
the [O IV] 25.89 $\mu$m 
line luminosity (\loiv, which is claimed to be
an intrinsic AGN luminosity indicator, see Mel\'{e}ndez et al. 2008a). 
The type II AGNs are shown with solid circles, while open circles represent 24 
type I AGNs for which the [O IV] line luminosity is available from literature.  
Solid stars denote tentative Compton-thick sources, based on previously 
reported measurements of N$_{\rm H}$ (e.g., Treister, Urry \& Virani 2009, and 
references therein). 
The right-hand panel shows the distribution of $L_{\rm Fe}$ for both type I (dotted line) and type II (solid line) 
AGNs.   
It can be seen that the distribution of the \fekalfa line luminosity 
in type II AGNs is not significantly different from that in type I AGNs. 
A K-S test shows the probability that both samples were drawn from the same parent population is 0.21.}
{ However, when normalizing the \fekalfa line luminosity by the [O IV] line luminosity ($L_{\rm Fe}$/\loiv, the lower panel of Figure 6), we find 
 a marginal difference in the distributions of
the $L_{\rm Fe}$/\loiv~ratio for type I and type II AGNs, but the significance level 
is not high (at 93\% confidence), possibly due to the small size of the current 
\chandra grating sample.
The dotted and dashed horizontal lines show the means of $L_{\rm Fe}$/\loiv~ for Liu \& Wang's type I and type II subsamples, respectively. 
 It seems that the distribution of the $L_{\rm Fe}$/\loiv~ratio for our \chandra grating 
sample shows a similar pattern as Liu \& Wang's sample.  
 However, we note that there is significant overlap between 
the type I and type II AGNs with intermediate values of the $L_{\rm Fe}$/\loiv~ratio 
(see also Liu \& Wang 2010, Figure 6), suggesting the complex 
dependencies of the \fekalfa line emission on the geometry and physical properties 
of the line-emitting material.  
In the context of a toroidal geometry (e.g., Murphy \& Yaqoob 2009), the sources in
the overlap region of the $L_{\rm Fe}$/\loiv~ plot may however indicate that they constitute a selected distribution of N$_H$ and inclination angle (e.g., Circinus galaxy, Yang et al. 2009), 
but a larger sample is required 
to make a detailed comparison.   
}

Although in our study we cannot determine some of the
critical parameters of the \fekalfa line emitter, such as global covering factor, $N_H$, and the inclination angle of 
the X-ray reprocessor,
we can still make interesting statements from the plot of \loiv~vs. $L_{\rm Fe}$/\loiv.
It is apparent that the ratio of $L_{\rm Fe}$/\loiv ~appears to have a dispersion of 
2 to 3 orders of magnitude or so.
Regardless of AGN type, the sources located at the top end of the $L_{\rm Fe}$/\loiv~ ratio distribution 
are expected to be those objects that are observed
at a face-on, or near face-on, inclination angle, with moderate 
$N_H$ ($\sim0.5-2\times10^{24}$ cm$^{-2}$). The sources located in the region of
the lowest values of $L_{\rm Fe}$/\loiv~should correspond to cases with higher $N_H$ and higher inclination angles.
We note that
the source with the lowest ratio of $L_{\rm Fe}$/\loiv~ is 
the prototypical Seyfert 2 galaxy NGC 1068, for which 
an edge-on orientation of the torus and a high, Compton-thick column density has
already been suggested in literature (e.g., Pounds et al. 2006, and references therein).

\section{SUMMARY}
\label{summary}

We have presented an empirical analysis of 29 observations of the
narrow core of the \fekalfa emission line in 10 type II AGNs 
using \chandra HEG data. 
The \fekalfa line was significantly detected, and its parameters (line centroid energy, intrinsic width and line flux) were
well-measured, in 8 out of 10 sources. 
The centroid energy of \fekalfa line is found to be strongly peaked around $\sim 6.4$~keV, indicating an
 origin in cool, neutral matter, consistent with the results of Paper I for a sample of type I AGNs. 

We obtained a weighted mean value of FWHM~$=2000 \pm 160\ \rm km \ s^{-1}$
for the intrinsic \fekalfa line width.
For five sources with spectro-polarimetric observations, we constructed 
99\% confidence, two-parameter contours of line flux versus the
ratio of the width of the \fekalfa line to the width of the H$\beta$ line.
We found that the
99\% confidence bounds on the ratio of the X-ray line width
to the optical line width lies in the range $\sim$0.3 -- 1.2, 
suggesting that contributions to the flux of the core of the \fekalfa line
are allowed from a region that is within a factor $\sim 0.7-11$ times the radius
of the optical BLR. 
Compared to 13 type I AGNs with sufficiently high quality \fekalfa line-width 
measurements (reported in paper~I), we found no difference in the distribution between the \fekalfa FWHM in 
type I and type II AGNs, and this conclusion is independent of the central black mass. 
This result suggests there may be a universal location for the bulk of the \fekalfa line emission 
with respect to the gravitational radius ($r_g$). 
However, these conclusions are subject to the caveat that derivation
of the true velocity width of the \fekalfa line core requires
a realistic physical model, and this will be the subject of future work.
 
Having isolated the narrow core of the \fekalfa line
with the best available spectral resolution, 
we also presented measurements of the line luminosity of the \fekalfa core, and 
examined its relation to the intrinsic AGN luminosity (i.e., 
by means of the \loiv~ indicator). { We found a marginal difference in 
the distribution of the
\fekalfa emission-line luminosity between type I and type II AGNs, 
but the significance level is not high, and the spread in the 
$L_{\rm Fe}$/\loiv ~is about 2 orders of magnitude.} 
Although the complex dependencies of the \fekalfa emission-line parameters
upon the covering factor, inclination angle and column density, 
prevent trivial use of the \fekalfa luminosity as a proxy of the intrinsic AGN's 
luminosity, the \chandra results presented here will provide an important 
and new supplement to additional X-ray spectroscopy with a broader bandpass.
A detailed comparison of the data for the \fekalfa line and the continuum with 
appropriate models will then yield more robust constraints on the intrinsic
AGN luminosity and the physical parameters of the X-ray processor.

{ We thank E. Moran for helpful discussions on the spectropolarimetric 
observations of NGC 4507.}  
X.W.S. thanks the support from China postdoctoral foundation. 
We acknowledge support from Chinese National Science Foundation 
(Grant No. 10825312), and the 
Fundamental Research Funds for the Central Universities 
(Grant No. WK2030220004, WK2030220005). This research
made use of the HEASARC online data archive services, supported
by NASA/GSFC. This research has made use of the NASA/IPAC Extragalactic Database
(NED) which is operated by the Jet Propulsion Laboratory, California Institute
of Technology, under contract with NASA.
The authors are grateful to the \chandra 
instrument and operations teams for making these observations
possible.

\newpage

\section*{Figure Caption}

\par\noindent
{\bf Figure 1} \\
\chandra HEG spectra in the Fe K band 
for sources in our sample. 
For eight AGN which were observed more than once, the time-averaged 
spectra are shown.  
The data are binned at $0.01\AA$, comparable to  
the HEG spectral resolution, which is $0.012\AA$
FWHM. The data are combined from the $-1$ and $+1$ orders
of the grating. The spectra have been corrected for instrumental effective area
and cosmological redshift.
Note that these are {\it not} unfolded spectra and are
therefore independent of any model that is fitted. 
The statistical errors shown correspond
to the $1\sigma$ 
(asymmetric) Poisson errors, which we calculated using 
equations (7) and (14) in Geherls (1986) that approximate the upper and
lower errors respectively.
The solid line corresponds to a continuum model fitted over the
3--10~keV range, 
as described in the text (Section 2).
The vertical dotted lines represent (from left to right), the rest energies
of the following: Fe~{\sc i}~$K\alpha$, \fexxv forbidden,
two intercombination lines of \fexxvp, \fexxv resonance, \feklyap,
Fe~{\sc i}~$K\beta$, and the Fe~K edge.

\begin{figure*}[h]
\vspace{10pt}
\centerline{\psfig{file=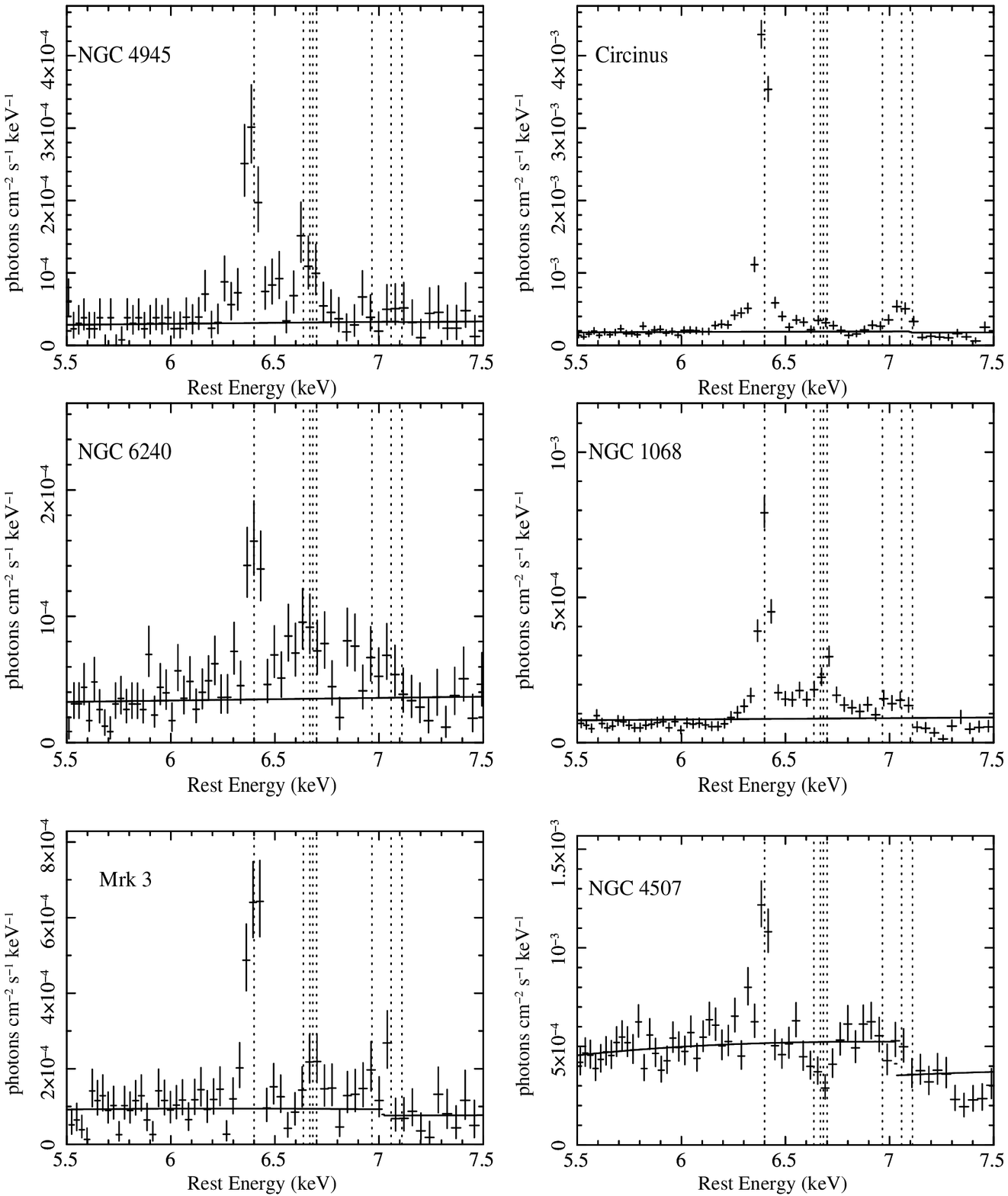,width=6.5in,height=8.in,angle=0}}
\caption{\footnotesize
}
\end{figure*}


\setcounter{figure}{0}
\begin{figure*}[tbh]
\vspace{10pt}
\centerline{\psfig{file=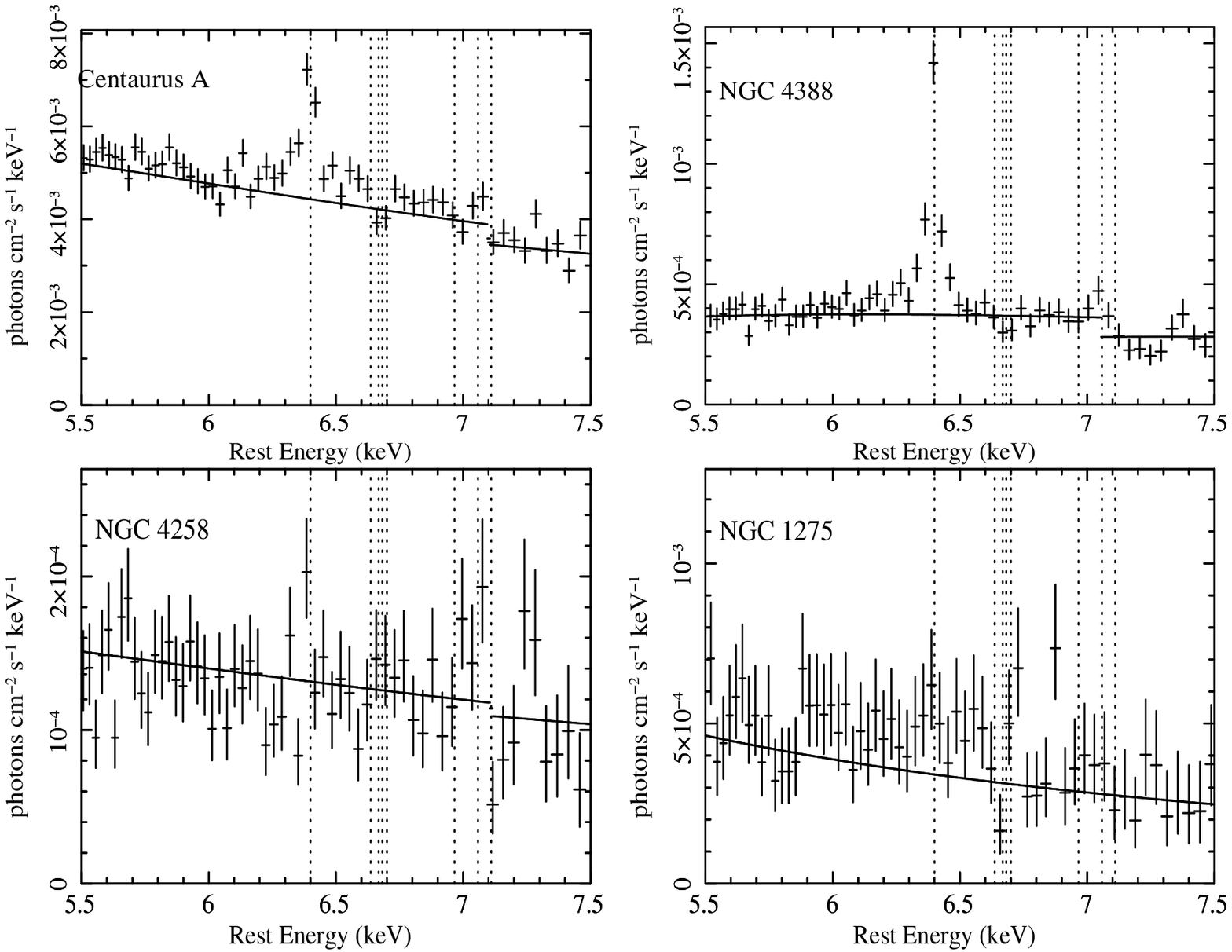,width=6.5in,height=5.5in,angle=0}}
\caption{ -- {\it continued}}
\end{figure*}


\begin{figure}
\epsscale{1.0}
\plotone{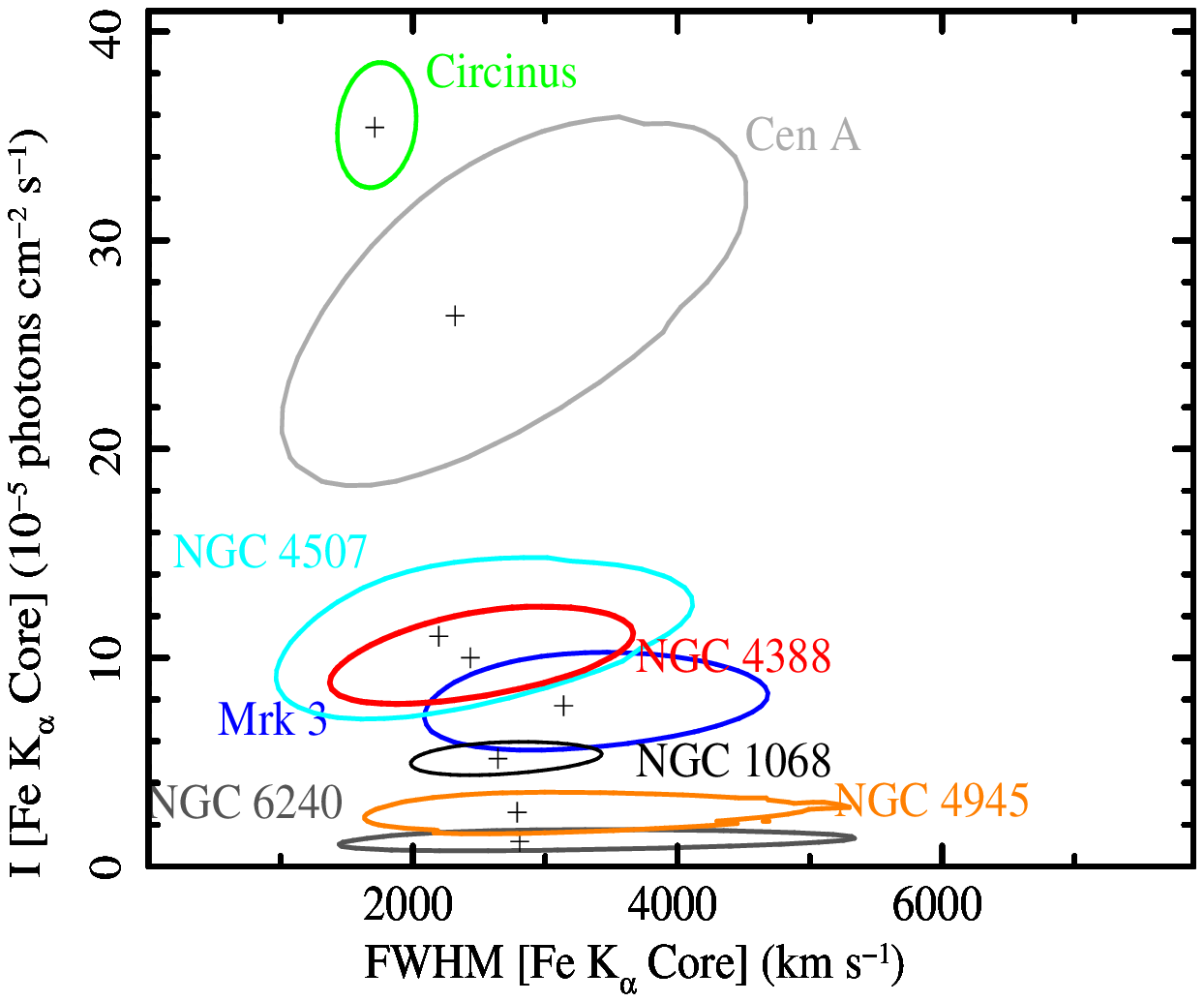}
\caption{
Joint 99\% confidence contours of the \fekalfa emission line intensity vs. velocity width 
(FWHM), obtained 
from Gaussian fits to the line as described in the text, for eight AGNs: NGC 4945 (orange), Circinus 
(green), NGC 6240 (light grey), NGC 1068 (black), Mrk 3 (blue), NGC 4507 (light blue), Centarus A 
(dark grey) and NGC 4388 (red).
The well-constrained contours (see also Fig. 3) suggest that our single-Gaussian fits are picking up an
intrinsic narrow component at $\sim$ 6.4 keV, and the effects of any complex continuum
components on the line parameters 
are small.  
}
 \end{figure}

\begin{figure}
\epsscale{1.0}
\plotone{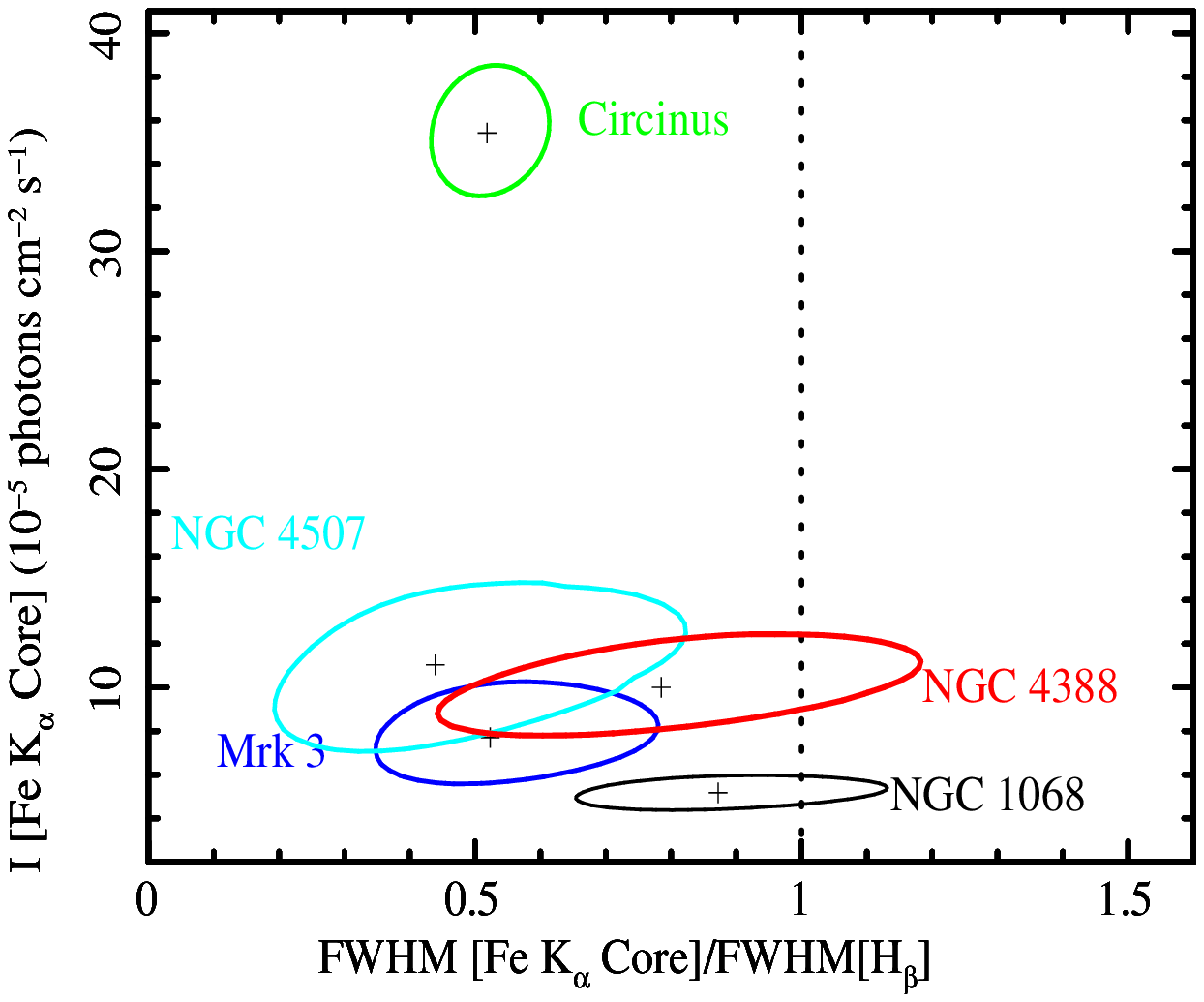}
\caption{
Joint 99\% confidence contours of the \fekalfa emission-line
core intensity versus the ratio of the \fekalfa FWHM to the H$\beta$ FWHM for
5 AGN that provided the values of  H$\beta$ line FWHM from spectro-polarimetric observation (see text).
For Circinus and NGC 4507, we used the FWHM of H$\alpha$ line as
a surrogate for H$\beta$ FWHM.
The vertical dotted lines correspond to a FWHM ratio
of the pairs of emission lines equal to unity.}
 \end{figure}

\begin{figure}
\epsscale{1.2}
\plotone{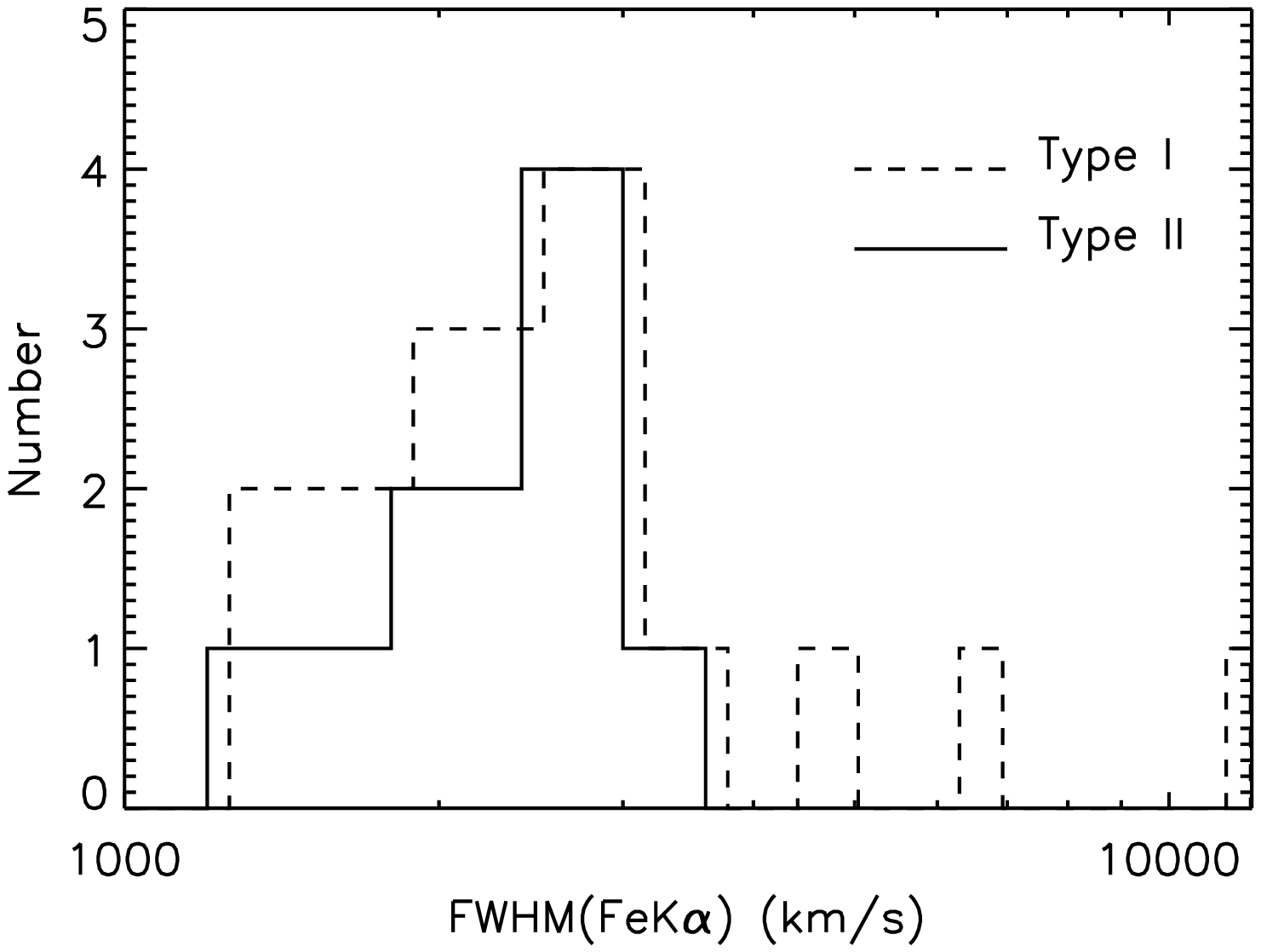}
\caption{
Distribution of the \fekalfa line FWHM 
derived from those sources for which the best \fekalfa line FWHM constraints
were obtained. Dashed and solid lines correspond to the distribution for
type~I (see Paper I) and type II AGNs, respectively.}
\end{figure}
\clearpage
\begin{figure}
\epsscale{1.5}
\centerline{\psfig{file=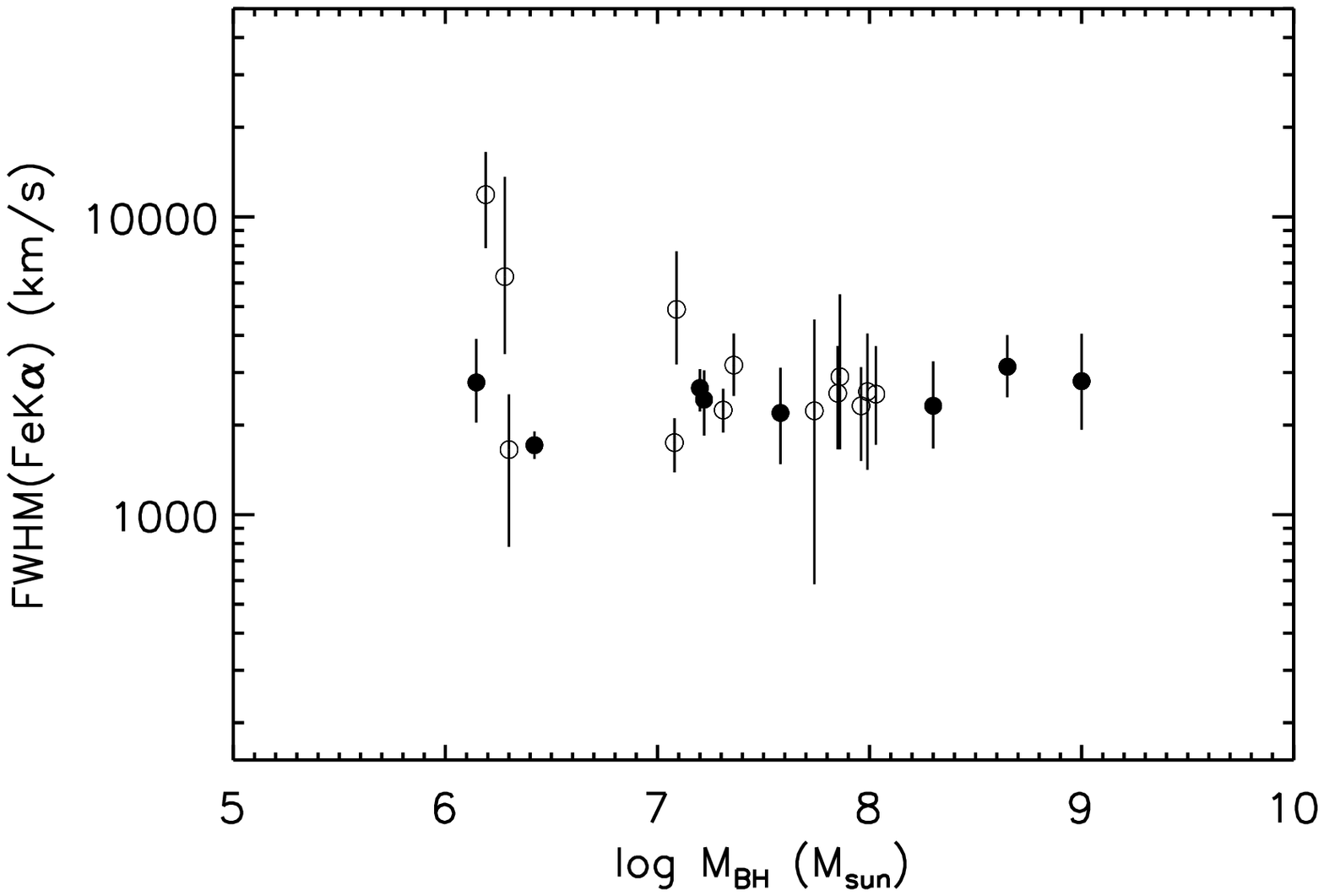,width=8.0in,height=5.5in,angle=0}}
\caption{\fekalfa emission-line FWHM versus the black hole mass. Solid circles denotes type II AGNs, and 
open circles correspond to the 13 type I sources shown in Figure 4. The statistical errors on the \fekalfa line FWHM shown correspond 
to 68\% confidence for three free parameters. It can be seen that 
while the black hole mass spans a range from $10^6$ to $10^9$ M$\odot$, the \fekalfa 
line FWHM remains nearly constant, clustering at $\sim$2000-3000 km s$^{-1}$ (see text).
\label{fig2}}
\end{figure}

\begin{figure}
\epsscale{1.5}
\centerline{\psfig{file=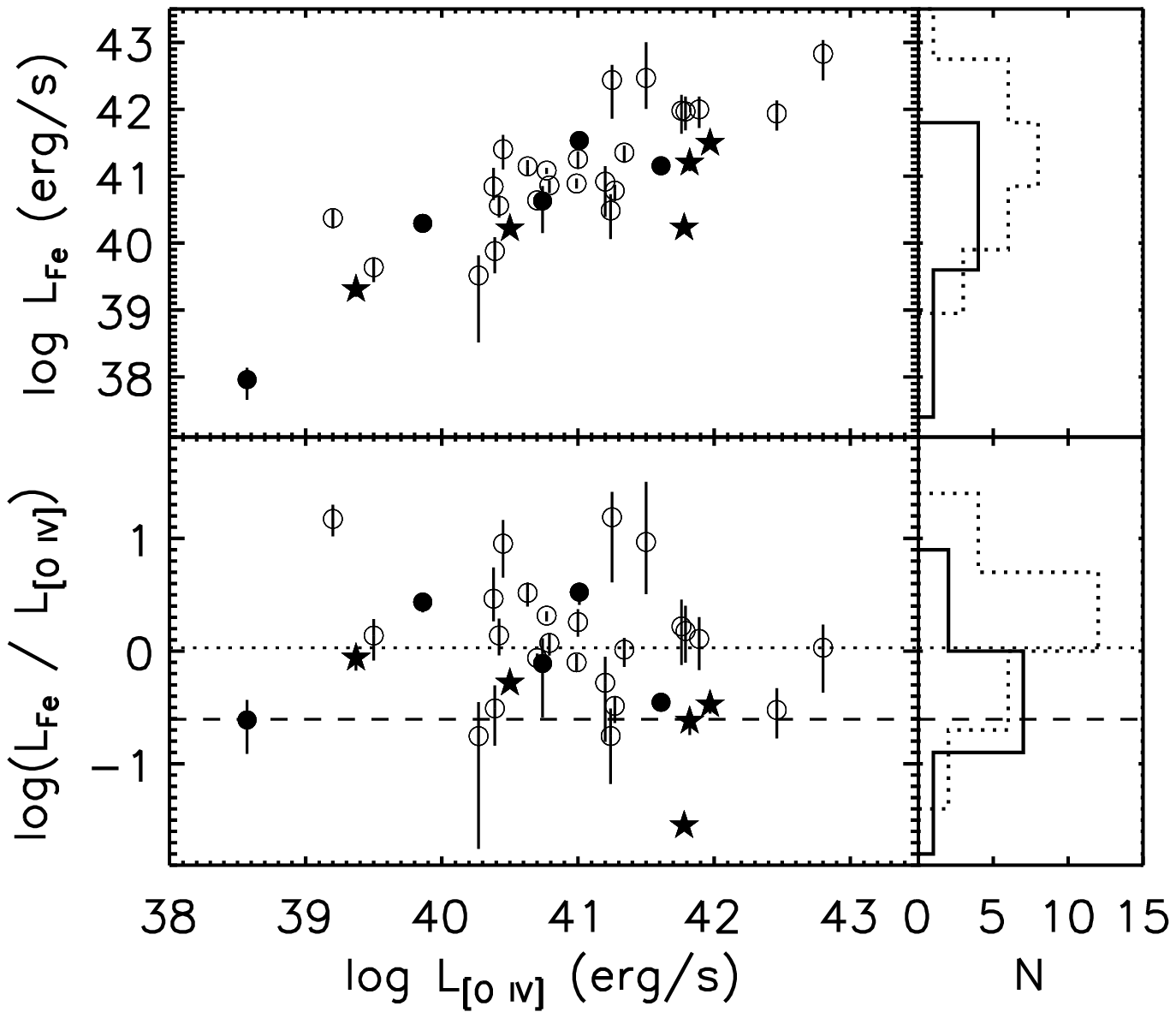,width=8.0in,height=5.5in,angle=0}}
\caption{
The [O IV] $\lambda25.89 \mu$m line luminosity versus the \fekalfa core emission-line luminosity 
(upper panel). The lower panel shows the
[O IV] $\lambda25.89 \mu$m line luminosity versus the ratio of \fekalfa luminosity to [O IV] luminosity.
Both solid and open circles have the same meaning as in Figure 5, while 
Compton-thick AGNs are distinguished by the solid stars. 
The right-hand panel is the distribution of the \fekalfa luminosity and the ratio of $L_{\rm Fe}/L_{[O IV]}$, 
respectively. 
The statistical errors on the \fekalfa line luminosity correspond to 68\% confidence ($\Delta C=3.506$, or 0.989, depending 
on whether there were three parameters or one parameter free respectively). 
The dotted and dashed horizontal lines represent the means of the $L_{\rm Fe}$/\loiv~ for 
Liu \& Wang (2010)'s type I and type II subsamples, respectively.
\label{fig2}}
\end{figure}

\newpage
\begin{deluxetable}{lccccc}
\tabletypesize{\scriptsize}
\tablecaption{\scshape Observation log of the \chandra HEG sample in this work}
\tablewidth{0pt}
\tablehead{\colhead{Source} & \colhead{$z$} &\colhead{SeqNum} & \colhead{ObsID}&  \colhead{Exposure (ks)} & \colhead{ObsDate}}
\startdata
 NGC 4945 & 0.001878 & 700981 & 4899 & 78.6  & 2004-05-28  \\
&                      & 700981 & 4900 & 95.8  & 2004-05-29  \\
Circinus & 0.001448 & 700046 & 374 & 7.3 & 2000-06-15 \\
         &          & 700268 & 62877 & 61.4 & 2004-06-02 \\
         &          & 700853 & 4770 & 56.1 & 2004-06-02 \\
         &          & 700854 & 4771 & 60.2 & 2004-11-28 \\    
NGC 6240 & 0.024480 & 701324 & 6908 & 159.0 & 	2006-05-16 \\
         &          & 701324 & 6909 & 143.0 & 2006-05-11 \\  
NGC 1068 & 0.003793 & 700004 & 332 & 46.3 & 2000-12-04 \\
         &          & 701591 & 9148 & 80.9 & 2008-12-05 \\
         &          & 701591 & 9149 & 91.2 & 2008-11-19  \\
         &          & 701591 & 10815 & 19.4 & 2008-11-20  \\
         &          & 701591 & 10816 & 16.4 & 2008-11-18 \\
         &          & 701591 & 10817 & 33.2 & 2008-11-22 \\ 
        &          & 701591 & 10829 & 39.1 & 2008-11-30 \\ 
       &          & 701591 & 10830 & 43.6 & 2008-12-03 \\ 
       &          & 701592 & 9150 & 41.8 & 2008-11-27 \\ 
        &          & 701592 & 10823 & 35.1 & 2008-11-25 \\
Mrk 3   & 0.013509         & 700178 & 873   & 101.9 & 2000-03-18 \\
NGC 4507 & 0.011801        & 700340 & 2150 & 139.8 & 2001-03-15 \\
Centaurus A & 0.001825     & 700216 & 1600  & 47.5 & 2001-05-09 \\
            &      & 700217 & 1601  & 52.2 & 2001-05-21 \\
NGC 4388 & 0.008419        & 701717 & 9276 & 172.8 & 2008-04-16 \\
         &         & 701717 & 9277 & 99.6 & 2008-04-24 \\
NGC 4258 & 0.001494 & 701543 & 7879 & 152.9 & 2007-10-08 \\
         &         & 701543 & 7880 & 60.0 & 2007-10-12 \\
         &         & 701543 & 9750 & 107.1 & 2007-10-14 \\
NGC 1275 & 0.017559 & 700005 & 333 & 27.4 & 1999-10-10 \\
         &         & 700201 & 428 & 25.0 & 2000-08-25 \\
\enddata
\end{deluxetable}

\newpage
\begin{deluxetable}{lccccccccc}
\tabletypesize{\scriptsize}
\tablecaption{\scshape Parameters of the Core \fekalfa Line Emission from $\it Chandra$ (HEG) Data}
\tablewidth{0pt}
\tablehead{\colhead{Source} & \colhead{$F_{\rm 2-10~keV}$} &\colhead{$E$} & \colhead{$EW$}&  \colhead{FWHM(\fekalfa)} 
&\colhead{I(\fekalfa)} & \colhead{L(\fekalfa)} & \colhead{FWHM (H$\beta$)} & \colhead{$M_{\rm BH}$} & \colhead{$L_{\rm [O IV]}$} \\
\hspace*{15.mm} &  &(keV)   &(eV) & (km s$^{-1}$) & & &(km s$^{-1}$)   & & \\
\hspace*{15.mm} (1)&(2) & (3) & (4) & (5) & (6) & (7) & (8)  & (9) & (10) }
\startdata
 NGC 4945(2)$^{\star}$ & 2.6 &$6.389_{-0.008}^{+0.007}$ & $840_{-191}^{+164}$ & 2780$_{-740}^{+1110}$& 
2.6$_{-0.6}^{+0.5}$ & 39.31$_{-0.11}^{+0.08}$ & $\dots$ & 6.2 & 39.37 \\
Circinus(4)$^\star$ & 16.0 &$6.396_{-0.001}^{+0.002}$ & $1673_{-83}^{+91}$ & 1710$_{-170}^{+190}$&
35.4$_{-1.8}^{+1.9}$ & 40.22$_{-0.02}^{+0.02}$  & 3300$^{\dag}$ & 6.1 & 40.50 \\
NGC 6240(2)$^\star$ & 2.9 &$6.394_{-0.009}^{+0.009}$ & $333_{-79}^{+96}$ & 2810$_{-880}^{+1240}$& 
1.2$_{-0.3}^{+0.3}$ & 41.20$_{-0.12}^{+0.10}$  & $\dots$ & 9.0 & 41.82 \\
NGC 1068(10)$^\star$  & 5.6 &$6.402_{-0.003}^{+0.003}$ &$779_{-76}^{+72}$ &2660$_{-440}^{+410}$&
5.2$_{-0.5}^{+0.5}$ & 40.23$_{-0.04}^{+0.04}$  & 3030 & 7.2 & 41.78 \\
Mrk 3$^\star$ & 6.3 & $6.396_{-0.008}^{+0.007}$ & 612$_{-104}^{+123}$& 3140$_{-660}^{+870}$&
7.7$_{-1.3}^{+1.5}$ & 41.50$_{-0.08}^{+0.08}$ & 6000 & 8.6 & 41.97 \\
NGC 4507 & 27.9 & $6.395_{-0.006}^{+0.007}$ & 114$_{-26}^{+21}$& 2200$_{-720}^{+910}$&
11.0$_{-2.5}^{+2.0}$  & 41.53$_{-0.11}^{+0.07}$  & 5000$^{\dag}$ & 7.6 & 41.01 \\
Centarus A(2) & 310.8 & $6.396_{-0.007}^{+0.005}$& 45$_{-8}^{+9}$ & 2320$_{-650}^{+950}$&
26.4$_{-5.0}^{+5.2}$ & 40.30$_{-0.09}^{+0.08}$ & $\dots$ & 8.3 & 39.86 \\
NGC 4388(2) &  23.1 & $6.393_{-0.004}^{+0.004}$& 169$_{-21}^{+24}$ & 2430$_{-590}^{+620}$&
10.0$_{-1.2}^{+1.4}$ & 41.16$_{-0.06}^{+0.06}$ & 3100 & 7.2 & 41.61\\
NGC 4258(3) & 9.0 & 6.4 (fixed) & 14$_{-9}^{+9}$ & 100 (fixed) & 
0.2$_{-0.1}^{+0.1}$  & 37.96$_{-0.30}^{+0.18}$ &  $\dots$  & 7.6 & 38.57 \\
NGC 1275(2) & 30.2 & 6.4 (fixed) & 19$_{-15}^{+14}$ & 100 (fixed) & 
0.6$_{-0.4}^{+0.4}$ & 40.63$_{-0.48}^{+0.22}$ & $\dots$  & 8.5 & 40.74 \\
\enddata
\tablecomments{
Results from
{\it Chandra } HEG data, fitted with a absorbed power law plus Gaussian emission-line model
in the 3--10 keV band. All parameters are quoted in the source rest
 frame. 
 Statistical errors are for the 68\% confidence
level for three free parameters in the Gaussian component of the model (corresponding to $\Delta$C=3.506). 
For NGC 4258 and NGC 1275, the 1$\sigma$ statistical errors are for one free parameter, corresponding to 
$\Delta$C=0.989. 
Col.(1): Source Name, whilst parentheses show the number of the observations 
 used to produce the time-averaged spectrum. 
{All HEG observations were in the \chandra public archives as of 2010 Aug 1 (see Section 2). } 
$^{\star}$Compton-thick AGN; 
{ Col.(2):  The observed 2--10 keV flux in units of 10$^{-12}$ ergs cm$^{-2}$ s$^{-1}$;} 
Col.(3): Gaussian line centroid energy;
Col.(4): Emission line equivalent width;
Col.(5): Full width half maximum of the \fekalfa, rounded to $10 \ \rm km \ s^{-1}$.
Col.(6): Emission-line intensity in units of 10$^{-5}$ photons cm$^{-2}$ s$^{-1}$.
Col.(7): The logarithm of \fekalfa line luminosity in units of  $\rm ergs\ s^{-1}$;
Col.(8): Full width half maximum of the broad polarized H$\beta$ line, 
 refers to Oliva et al. 1998; 
Nishiura \& Taniguchi 1998; Moran et al. 2000; Young et al. 1996. $^{\dagger}$ Broad polarized H$\alpha$ line. 
Col.(9): Black hole mass, refers to: Greenhill, Moran, \& Herrnstein (1997); Bian \& Gu (2007); 
Tecza et al. (2000); Tremaine et al. (2002); Marconi et al. (2001); Woo \& Urry (2002). 
Col.(10): Logarithm of the [O IV] emission line luminosity, refers to: Liu \& Wang (2010); 
Mel\'{e}ndez et al. (2008b).} 
\end{deluxetable}


\begin{thebibliography}{59}

\bibitem{} Antonucci, R. 1993, ARA\&A, 31, 473 

\bibitem{arnaud1996} Arnaud, K. A., 1996, Astronomical Data Analysis
Software and Systems V, eds. Jacoby, G., \& Barnes, J., p. 17,
ASP Conference Series, Vol. 101

\bibitem{} Bian W., \& Gu Q. 2007, \apj, 657, 159

\bibitem{} Bianchi S., Matt, G., \& Fiore, F. et al. 2002, A\&A, 396, 793

\bibitem{} Bianchi S., La Franca F., 
Matt G., Guainazzi M., Jim\'{e}nez-Bail\'{o}n E., 
Longinotti A. L., Nicastro F., Pentericci L., 2008, MNRAS, 389, 52


\bibitem{} Bianchi, S., Piconcelli, E., Chiaberge, M. et al. 2009, \apj, 695, 781

\bibitem{brightman2010}
Brightman M., \& Nandra K., 2011, MNRAS, in press (arXiv:1012.3345)

\bibitem{cai2010} Cai, H. B., Shu, X. W., Wang, J. X., \& Zheng, Z. Y. 2010, RAA, 10, 427 

\bibitem{c01} Conselice, C. J., Gallagher, J. S., Wyse, R. F. G. 2001, \aj, 122, 2281 

\bibitem{e04} Elvis, M., Risaliti, G., Nicastro, F. et al. 2004, ApJ, 615, 25

\bibitem{f10} Fukazawa, Y. Hiragi, K., Mizuno, M. et al. 2011, \apj, 727, 19

\bibitem{gaskell2008} Gaskell C. M., Goosmann R. W., Klimek E. S., 2008, MmSAI, 79, 1090

\bibitem{} Gehrels, N. 1986, \apj, 303, 33
%
\bibitem{} Greenhill L. J., Moran J. M., Herrnstein J. R. et al. 1997, \apj, 481, L23

\bibitem{ikeda2009} Ikeda S., Awaki, H., \& Terashima Y. 2009, \apj, 692, 608

\bibitem{Jiang06} Jiang, P., Wang, J. X., \& Wang, T. G. 2006, \apj, 644, 725

\bibitem{lan08} Landt, H., Bentz, M. C., Ward, M. J. et al. 2008, ApJS. 174, 282

\bibitem{lee02} Lee J. C., Iwasawa K., Houck J. C. et al. 2002, \apj, 570, L47

\bibitem{levenson2002} Levenson N. A., Krolik J. H., Zycki, P. Y. et al. 2002, \apj, L81

\bibitem{levenson2006} Levenson N. A., Heckman T. M., Krolik J. H.,
Weaver K. A., \.{Z}ycki P. T., 2006, ApJ, 648, 111

\bibitem{liu2010} Liu Y., Elvis, M., McHardy, I. M. et al. 2010, \apj, 710, 1228

\bibitem{liu2010} Liu T., \& Wang, J. X. 2010, \apj, 725, 238


\bibitem{lz2001} Lubi\'{n}ski P., Zdziarski A. A., 2001, MNRAS, 323, L37 

\bibitem{} Marconi, A., Capetti, A., Axon, D. J. et al. 2001, \apj, 549, 915

\bibitem[Markert \etal 1995]{mark1995} Markert T. H., 
Canizares C. R., Dewey D.,
McGuirk M., Pak C., Shattenburg M. L., 1995, Proc. SPIE, 2280, 168

\bibitem{mat02} Matt, G. 2002, \mnras, 337, 147

\bibitem{} Mel\'{e}ndez, M., et al. 2008a, ApJ, 682, 94 

\bibitem{} Mel\'{e}ndez, M., et al. 2008b, ApJ, 689, 95

\bibitem{miller08} Miller L., Turner T. J. \& Reeves J. N. 2008, A\&A, 483, 437

\bibitem{} Moran E. C., Barth A. J., Kay L. E., \& Filippenko A. V. 2000, ApJ, 540, L73

\bibitem{murphy2009} Murphy  K., Yaqoob T., 2009, MNRAS, 397, 1549 
 
\bibitem{Nand2006} Nandra  K., 2006, MNRAS, 368, L62

\bibitem{Netzer90} Netzer, H., et al. 1990, ApJ, 353, 108 

\bibitem{} Nishiura S., \& Taniguchi Y. 1998, \apj, 499, 134

\bibitem{} Oliva E., Marconi A., Cimatti A., Alighieri S. D. 1998, A\&A. 329, 21

\bibitem{} Palmeri P., Mendoza C., Kallman T. R., 
Bautista M. A., Melendez M., 2003, A\&A, 410, 359

\bibitem{Perola2002} Perola G. C., Matt G., Cappi M., Fiore F.,
Guainazzi M., Maraschi L., Petrucci P. O., Piro, L., 2002, A\&A, 389, 802

\bibitem{Peterson04} Peterson B. M., Ferrarese L., Gilbert K. M. et al. 2004, \apj, 613, 682

\bibitem{} Pike C. D., Phillips K. J. H., Lang J.
\etal 1996, ApJ, 464, 487

\bibitem{} Pounds K., \& Vaughan S. 2006, \mnras, 368, 707


\bibitem{} Ramos Ameida, C. et al. 2008, \apj, 680, L17

\bibitem{Rey09} Reynolds C. S., Nowak M. A., \& Markoff S. et al. 2009, \apj, 691, 1159

\bibitem{r05} Risaliti, G., Elvis, M., Fabbiano, G., Baldi, A.,\& Zezas, A. 2005, \apj, 623, L93


\bibitem{Shu10} Shu  X. W., Yaqoob T. \& Wang J. X. 2010, \apjs, 187, 581 (Paper I)

\bibitem{Sulentic1998} Sulentic J. W., Marziani P., Zwitter T.,
Calvani M., Dultzin-Hacyan, D., 1998, ApJ, 501, 54

\bibitem{} Tecza M., Genzel R., Tacconi L. J. et al. 2000, \apj, 537, 178

\bibitem{} Treister E., Urry C. M., \& Virani S. 2009, \apj, 696, 110

\bibitem{} Tremaine S., Gebhardt K., Bender R. et al. 2002, \apj, 574, 740

\bibitem{} Verner D. A., Ferland G. J., Korista K. T., Yakovlev D. G., 1996, ApJ, 465, 487

\bibitem{Wgyy2001} Weaver K. A., Gelbord J., Yaqoob T., 2001, ApJ, 550, 261

\bibitem{Winter2009} Winter L. M., Mushotzky R. F., Reynolds C. S., Tueller J.,
 2009, ApJ, 690, 1322

\bibitem{} Woo J. H., \& Urry C. M. 2002, \apj, 579, 530

\bibitem{Yamada09} Yamada S., Itoh T., Makishima K., \& Nakazawa, K. 2009, PASJ, 61, 309

\bibitem{Yang09} Yang Y., Wilson A. S., Matt G., Terashima Y., Greenhill L. J. 2009, \apj, 691, 131

\bibitem{Yaq2001} Yaqoob T., George I. M., Nandra K., Turner T. J., 
Serlemitsos P. J., Mushotzky R. F., 2001, ApJ, 546, 759

\bibitem{Yaq2003} Yaqoob T., George I. M., Kallman T. R.,
Padmanabhan U., Weaver. K. A., 
Turner T. J., 2003, ApJ, 596, 85 

\bibitem{} Yaqoob T., Padmanabhan U., 2004, ApJ, 604, 63 (YP04)

\bibitem{yaq2007} Yaqoob T., Murphy K. D., Griffiths R. E. et al. 2007, PASJ, 59, 283
 
\bibitem{yaq2010a} Yaqoob T., \& Murphy, K. D. 2011, \mnras, 412, 277

\bibitem{yaq2010b} Yaqoob T., \& Murphy, K. D., Miller, L. \& Turner, T. J. 2010, \mnras, 401, 411

\bibitem{} Young S. et al. 1996, \mnras, 281, 1206

\bibitem{young2005} Young  A. J., Lee  J. C., Fabian  A. C.,
Reynolds  C. S., Gibson  R. R., Canizares C. R., 2005, ApJ, 631, 733



\end{thebibliography}
\end{document}